\documentclass[conference]{IEEEtran}
\IEEEoverridecommandlockouts
\usepackage{cite}
\usepackage{amsmath,amssymb,amsfonts}
\usepackage{algorithmic}
\usepackage{graphicx}
\usepackage{textcomp}
\usepackage{xcolor}
\def\BibTeX{{\rm B\kern-.05em{\sc i\kern-.025em b}\kern-.08em
    T\kern-.1667em\lower.7ex\hbox{E}\kern-.125emX}}

\newcommand{\bQ}{{\mathbf Q}}
\newcommand{\cA}{{\mathcal A}}
\newcommand{\cN}{{\mathcal N}}

\newcommand{\cP}{{\mathcal P}}
\newcommand{\cR}{{\mathcal R}}
\newcommand{\cS}{{\mathcal S}}
\newcommand{\dC}{{\mathbb C}}
\newcommand{\dE}{{\mathbb E}}
\newcommand{\dR}{{\mathbb R}}

\newcommand{\mI}{\mbox{$\bf I $}}

\newcommand{\mo}{\mbox{$\bf O$}}
\newcommand{\va}{\mbox{$\bf a $}}

\newcommand{\vc}{\mbox{$\bf c $}}
\newcommand{\vg}{\mbox{$\bf g $}}
\newcommand{\vh}{\mbox{$\bf h $}}

\newcommand{\vo}{\mbox{$\bf o $}}

\newcommand{\vs}{\mbox{$\bf s $}}

\newcommand{\vw}{\mbox{$\bf w $}}
\newcommand{\vx}{\mbox{$\bf x $}}

\newcommand{\be}{\begin{equation}}
\newcommand{\ee}{\end{equation}}
\newcommand{\bea}{\begin{eqnarray}}
\newcommand{\eea}{\end{eqnarray}}

\newcommand{\MYfooter}{\smash{
\hfil\parbox[t][\height][t]{\textwidth}{\centering
\thepage}\hfil\hbox{}}}

\makeatletter

\def\ps@headings{%
\def\@oddhead{\parbox[t][\height][t]{\textwidth}{\centering
Accepted for presentation at the IEEE GLOBECOM 2021, SAC, Machine Learning for Communications, \textcopyright 2021 IEEE
}}%
}

\def\ps@IEEEtitlepagestyle{%
\def\@oddhead{\parbox[t][\height][t]{\textwidth}{\centering
Accepted for presentation at the IEEE GLOBECOM 2021, SAC, Machine Learning for Communications, \textcopyright 2021 IEEE
}\hfil\hbox{}}%
\def\@evenhead{\scriptsize\thepage \hfil \leftmark\mbox{}}%
\def\@evenfoot{\MYfooter}}

\makeatother
\begin{document}

\title{Multi-agent deep reinforcement learning (MADRL) meets multi-user MIMO systems}
\author{\IEEEauthorblockN{Heunchul Lee}
\IEEEauthorblockA{\textit{Ericsson Research,} \\
\textit{Ericsson AB}\\
Stockholm, Sweden \\
heunchul.lee@ericsson.com}
\and
\IEEEauthorblockN{Jaeseong Jeong}
\IEEEauthorblockA{\textit{Ericsson Research,} \\
\textit{Ericsson AB}\\
Stockholm, Sweden \\
jaeseong.jeong@ericsson.com}
}

\maketitle

\begin{abstract}
A multi-agent deep reinforcement learning (MADRL) is a promising approach to challenging problems in wireless environments involving multiple decision-makers (or actors) with high-dimensional continuous action space. 
In this paper, we present a MADRL-based approach that can jointly optimize precoders to achieve the outer-boundary, called pareto-boundary, of the achievable rate region for a multiple-input single-output (MISO) interference channel (IFC).
In order to address two main challenges, namely, multiple actors (or agents) with partial observability and multi-dimensional continuous action space in MISO IFC setup,
we adopt a multi-agent deep deterministic policy gradient (MA-DDPG) framework in which decentralized actors with partial observability can learn a multi-dimensional continuous policy in a centralized manner with the aid of shared critic with global information.
Meanwhile, we will also address a phase ambiguity issue with the conventional complex baseband representation of signals widely used in radio communications.
In order to mitigate the impact of phase ambiguity on training performance, we propose a training method, called phase ambiguity elimination (PAE), that leads to faster learning and better performance of MA-DDPG in wireless communication systems. 
The simulation results exhibit that MA-DDPG is capable of learning a near-optimal precoding strategy in a MISO IFC environment.

\end{abstract}


\section{Introduction}
\subsection{Multi-cell MIMO problems and multi-agent system}
As cellular data demand continues to rise, an ultra-dense network is widely considered as a key component in managing this rising.
Multiple-input multiple-output (MIMO) technique has been developed for efficient transmission and reception of radio signals in multiple antenna systems. 
In particular, downlink multi-user MIMO is a promising technique to achieve higher throughput in a multi-cell environment. However, in general, the optimization problems in a multi-cell multi-user MIMO system are nonconvex and difficult to solve using the traditional approach based on mathematical models.
Machine learning (ML) is a promising approach to overcome the limitations of the traditional model-based approach, allowing the future cellular networks to evolve towards more scalable and intelligent architectures \cite{Wikstroem:20}.
In this paper, by leveraging the recent success of deep reinforcement learning (DRL) and multi-agent (MA) learning \cite{Lillicrap:16}\cite{Ryan:17}\cite{Foerster:18},
we propose a ML-based approach that integrates multi-agent deep reinforcement learning (MADRL) into downlink multi-cell multi-user wireless systems.

Ideally, joint data transmission schemes assume full MIMO cooperation in a multi-cell multi-user environment. 
For instance, coordinated multi-point (CoMP) with joint transmission (JT) is a cellular data transmission technique involving simultaneous transmission from multiple base stations (BSs) to the same user \cite{Dahlman:16}.
However, potential solutions to the JT scheme require significant amounts of global channel state information (CSI) and data sharing between the base stations, which is not only expensive but also difficult  in real-world cellular systems.
The optimal JT scheme can be reduced to coordinated beamforming (CB) schemes based on the transmission of the signal by a single base station that require local CSI and no inter-cell data sharing \cite{Dahlman:16}. In this case, each BS operates independently by treating the interference as background noise and the multi-cell multi-user setup can be modeled as MIMO interference channel (IFC). Compared to a single-cell system, the performance of MIMO IFC can be severely impacted by inter-cell interference, which becomes a crucial limiting factor. 

Reinforcement learning (RL) allows an agent to learn the optimal action policy that returns the maximum reward through trial-and-error interactions with a challenging dynamic environment \cite{Sutton:17}.
RL has been used to solve challenging problems in various areas ranging from games to robotics.
In wireless, RL is also emerging as one of key enablers for designing 6G AI-driven PHY-layer \cite{Wikstroem:20}. 
Recently, we have investigated RL-based approaches to improve the performance of MIMO systems \cite{lee:20}\cite{lee:20b}. 
However, these studies have  focused on enhancing the performance of single-cell MIMO systems.
Multi-cell multi-user precoding problems can be seen as a multi-agent system that learns to coordinate transmission schemes (or action policies) in interaction with other base stations (or other agents). 
Therefore, scaling our previous work to more complex multi-agent problems is crucial to building future intelligent networks that can operate in real-world multi-cell environments. 

The multi-agent problem requires complex inter-cell interference coordination in the sense that each BS should exhibit cooperative behavior to maximize the signal power to a desired user while minimizing the interference power to other users in the multi-cell environment.
We note that this problem setup poses two main challenges: i) {\it multiple actors (or agents) with partial observability} and ii) {\it multi-dimensional continuous action space}. 
The first challenge is a direct result of practical limitations of accessible information by local agents distributed in MISO-IFC,
and 
the second challenge comes from the fact that multi-dimensional precoding vectors should be optimized for multi-antenna BSs based on a certain transmit power constraint. 

\subsection{Main contributions}
To address these two challenges, we propose a multi-agent deep deterministic policy gradient (MA-DDPG)-based approach 
that can learn an optimal precoding strategy in multi-cell multi-user MIMO systems under the assumptions of local CSI and no inter-cell data sharing.
In particular, in order to permit tractable performance analysis, we consider a multiple-input single-output (MISO) IFC in which two base stations equipped with multiple antennas serve two single-antenna users, making the precoding problem tractable by the numerical methods proposed in \cite{Jorswieck:08} and \cite{Park:2010}. 
In this two-user MISO IFC setup, we can obtain the achievable rate region by using the work of \cite{Jorswieck:08} and derive the pareto-boundary of the rate region. 

MA-DDPG algorithms provide a multi-agent framework to learn a high dimensional continuous policy \cite{Ryan:17}\cite{Foerster:18}.
The actor-critic based policy gradient algorithms allow centralized training with decentralized execution in which local actors with partial observability can learn a globally optimal policy with the aid of centralized critic with global information at training time and execute the learned policy based only on partial observations at execution time. At the same time, the deterministic policy gradient (DPG) algorithm enables the agents to learn a multi-dimensional continuous policy.
The MA-DDPG framework is adopted to improve the quality of the received signals in MISO IFC by alleviating the inter-cell interference. 
Meanwhile, we also investigate the impact of phase ambiguity with the baseband representation of wireless channel on training performance.
The complex-valued representation of channel states has inherent phase ambiguity in the sense that the phase-shifted versions of a channel state will have the same impact in system performance as the original channel state.
From the wireless system design point of view, this phase ambiguity should not be a problem
but it can cause a performance degradation in a multi-agent learning system. 
In order to mitigate the impact of this phase ambiguity in training time, we propose a feature engineering method, called {\it phase ambiguity elimination (PAE)}, as a pre-processing step on input channel states to using a MA-DDPG algorithm to learn an optimal policy.
By applying the proposed PAE method in the state space, we demonstrate faster learning and better performance of MA-DDPG in MISO IFC. 
The simulation results indicate that MA-DDPG is capable of learning precoding schemes which achieves the outer boundary, called pareto-boundary, of achievable rate regions on MISO IFC environments.
To the best of our knowledge, this is the first work to demonstrate that the MA-DDPG framework can jointly optimize precoders to achieve the pareto-boundary of achievable rate region in a multi-cell multi-user multi-antenna system.

\section{System model} \label{sec:sys}
In this section, we describe the system model of MISO IFC, where two BSs equipped with multiple antennas simultaneously communicate with its own desired user equipment (UE) equipped with a single antenna in the same time-frequency resource.
It is important to recall that we have assumed this MISO IFC setup to ensure that the rate region can be obtained by the numerical method in \cite{Jorswieck:08}.
We also describe the numerical method for obtaining achievable rate region as well as two pareto-optimal rate pairs with closed-form expressions.
In the simulation section,  numerical results will be used as a quantitative criterion for demonstrating the optimality of MA-DDPG in multi-cell multi-user MISO systems.
\subsection{MISO IFC scenario}
In this subsection, we present a MISO IFC model and related assumptions. 
As shown in Figure  \ref{fmaddpg}, 
BS $i \in \{1,2\}$ desires to send the data symbol $d_i$ to UE $i$.
The base stations employ $n_t$ transmit antennas and each UE is equipped with a single receive antenna. 
BS $i$ employs a linear precoding vector $\vw_i$ of size $n_t$-by-1 prior to transmission over the air, which transforms the data symbol $d_i$ to the $n_t$-by-1 transmitted vector $\vx_i=\vw_i d_i$. 
The channel model from the BS $i$ to the two UEs are represented by an $1$-by-$n_t$ channel
vector $\vh_i=[h_{i,1},h_{i,2},\cdots,h_{i,n_t}]$ and $\vg_i=[g_{i,1},g_{i,2},\cdots,g_{i,n_t}]$, where the $j$-th elements $h_{i,j}$ and $g_{i,j}$ denote the path
gain from the $j$-th antenna of BS $i$ to the desired UE $i$ and the other UE, respectively. The channel elements
are independently and identically distributed (i.i.d.) according
to $\cN_{\dC}(0, 1)$, i.e., $\vh_i \in \dC^{n_t}$ and $\vg_i \in \dC^{n_t}$.

The received signal $y_i$ at UE $i$ can be expressed as, for $i=1$,
\bea \label{eq:rxsig1}
y_1=\vh_1\vx_1 +\vg_2\vx_2 + n_1=\vh_1\vw_1 d_1 +\vg_2\vw_2 d_2 + n_1,
\eea
and, for $i=2$,
\bea \label{eq:rxsig2}
y_2=\vh_2\vx_2+\vg_1\vx_1 + n_2=\vh_2\vw_2 d_2 +\vg_1\vw_1 d_1 + n_2,
\eea
where $n_i$ denotes complex-valued additive white Gaussian noise (AWGN) at  UE $i$, distributed as $\cN_{\dC}(0, \sigma_n^2)$.
Note that UE $i$ not only receives its desired signal $\vw_i d_i$ through the channel $\vh_i$ but also the inter-cell interference (ICI) $\vw_j s_j$ from the signal intended for the other UE $j\ne i$.
We impose the power constraint $\dE\left[tr[\vx_i(\vx_i)^H] \right]=1$ under assumption of unit-norm weight vectors $\vw_i$ and unit-power symbols $d_i$, i.e., $||\vw_i||=1$ and $\dE\left[|d_i|\right]=\sigma_d^2=1$, 
where $\dE\left[\cdot\right]$ denotes the expectation with respect to the distribution of the underlying random variable, 
$tr[\cdot]$ denotes the trace operator of a matrix,
$||\cdot||$ indicates the 2-norm of a vector,
and $\left|\cdot\right|$ denotes the absolute value of a scalar. 
Then the average transmit signal-to-noise ratio (SNR) of the network is defined as $\rho=\frac{1}{\sigma_n^2}$.

In the multi-cell environment, the signal quality is measured in the form of achievable data rate as a function of received signal-to-interference-plus-noise ratio (SINR).
From \eqref{eq:rxsig1}, the received SINR for UE $1$ is defined as
\bea \label{eq:sinr1}
SINR_1=\frac{\sigma_d^2\left|\vh_1\vw_1\right|^2}{\sigma_n^2+\sigma_d^2\left|\vg_2\vw_2\right|^2},
\eea

Similarly, the received SINR at UE $2$ is given by
\bea
SINR_2=\frac{\sigma_d^2\left|\vh_2\vw_2\right|^2}{\sigma_n^2+\sigma_d^2\left|\vg_1\vw_1\right|^2}.
\eea

\subsection{Rate region and two pareto-optimal rate pairs with closed-form expressions}
Let $r_i$ denote the rate for UE $i$. We denote by $R(\vw_1,\vw_2)$ the conditional rate tuple $(r_1,r_2)$ that can be achieved for a given pair of linear precoding vectors $\vw_1$ and $\vw_2$.
The theoretical limit of rate tuple $R(\vw_1,\vw_2)$ achievable with Gaussian random coding is given as 
\bea \label{eq:achi}
R(\vw_1,\vw_2)=\left[\log_2\left(1+\frac{\sigma_d^2\left|\vh_1\vw_1\right|^2}{\sigma_n^2+\sigma_d^2\left|\vg_2\vw_2\right|^2}\right) \right., \cr
\left. \log_2\left(1+\frac{\sigma_d^2\left|\vh_2\vw_2\right|^2}{\sigma_n^2+\sigma_d^2\left|\vg_1\vw_1\right|^2}\right)\right]. 
\eea
Then, the achievable rate region $\cR$ can be defined as the closure of the set of all achievable rate pairs $R(\vw_1,\vw_2)$ under the power constraints $||\vw_i||^2\leq 1$ for $i=1,2,$ 
\bea \label{eq:rr}
\cR =\cup_{||\vw_i||^2\leq 1,i=1,2}R(\vw_1,\vw_2).
\eea

Our goal is to find a pareto-optimal linear precoding scheme to construct pairs of precoding vectors $\vw_1$ and $\vw_2$ that achieve all the rate pairs on the pareto-boundary of the achievable rate region $\cR$. In general, the rate region is not known. But, we can numerically obtain the pareto-boundary by using the work in \cite{Jorswieck:08}. As shown in \cite{Jorswieck:08}, any pareto-optimal rate pairs can be achieved by using precoding vectors $\vw_1$ and $\vw_2$ that are parameterized by maximum ratio transmission (MRT) and zero-forcing (ZF) solution. 
The MRT and ZF solution at BS $i$ are given by
\bea \label{eq:mrti}
\vw_i^{mrt} =\vh_i^H.
\eea
and
\bea \label{eq:zfi}
\vw_i^{zf} = \left(\vg_i^H\vg_i\right)^{-1} \vh_i^H.
\eea
The MRT solution $\vw_i^{mrt}$ is optimal for single-user MIMO by maximizing the signal gain at the intended UE $i$.
In comparison, the ZF solution $\vw_i^{zf}$ can be seen as a MRT precoding vector designed for projecting the symbol $d_i$ on the null space of $\vg_i$.

As shown in Figure \ref{fsample}, we can numerically evaluate the rate region $\cR$ in \eqref{eq:rr} by using the parameterized precoding vectors $\vw_1$ and $\vw_2$ presented in \cite{Jorswieck:08}, and determine the pareto-boundary for two-user MISO IFC. 

We now briefly present two pareto-optimal rate pairs with closed-form expressions. 
The first reference rate pair can be directly achieved by applying the mixed pairs of two precoding vectors given in \eqref{eq:mrti} and \eqref{eq:zfi}, resulting in two rate pairs $R(\vw_1^{mrt},\vw_2^{zf})$ and $R(\vw_1^{zf},\vw_2^{mrt})$. As will be seen later in Figure \ref{frateregion}, the two rate pairs correspond to two corner points on the pareto-boundary.

The second reference rate pair is given by the leakage-based precoder scheme proposed in \cite{Sadek:07} that can maximize the received SINR at the receiver by maximizing the signal-to-leakage-and-noise ratio (SLNR) at the transmitter.

The SLNR at the UE $1$ is defined as
\bea 
SLNR_1=\frac{\sigma_d^2\left|\vh_1\vw_1\right|^2}{\sigma_n^2+\sigma_d^2\left|\vg_1\vw_1\right|^2}, 
\eea
where the second term on the denominator $\vg_1\vw_1$ indicates the {\it leakage} caused by the signal intended for the desired UE 1 to the other UE 2. 

The optimal SLNR solution can be obtained by
\bea 
\vw_1^{slnr} = \text{max-eigenvector}\left(\left(\sigma_n^2\mI+\vg_1^H\vg_1\right)^{-1} \vh_1^H\vh_1\right).
\eea
Similarly, the optimal SLNR solution at UE 2 is given by 
\bea
\vw_2^{slnr} = \text{max-eigenvector}\left(\left(\sigma_n^2\mI+\vg_2^H\vg_2\right)^{-1} \vh_2^H\vh_2\right).
\eea

The SLNR solution is known to achieve a sum-rate point on the pareto-boundary, corresponding to the rate pair in rate region that obtains the maximum sum rate. 
This sum-rate optimal rate pair, denoted by $R(\vw_1^{slnr},\vw_2^{slnr})$, as well as the two corner points $R(\vw_1^{mrt},\vw_2^{zf})$ and $R(\vw_1^{zf},\vw_2^{mrt})$ will serve as an upper reference rate pair  when we provide simulation results of MA-DDPG in the simulation section.

\section{Pareto-optimal precoding strategy} \label{sec:proposed}
In this section, we present a pareto-optimal precoding strategy based on MA-DDPG framework for a MISO IFC setup.
We first describe how MA-DDPG framework can be adopted to learn an optimal precoding strategy in MISO IFC setup. Then, we address the phase ambiguity issue with the conventional complex baseband representation in radio communications. In order to avoid the impact of this phase ambiguity in the learning process,  we propose a training method that leads to faster learning and better performance of MA-DDPG in wireless communication systems.

\subsection{Multi-agent RL with a continuous action space}
RL problems can be formalized by modelling the interaction between the agent and the environment as a \emph{Markov decision process} (MDP). An MDP consists of a set of environment states $\cS$, a set of available actions $\cA$, a reward $ r\in \dR$ and a state transition function $\cP: \cS \times \cA \rightarrow \cS$ from one state to another given an action taken.
A policy is a mapping function from state to action in the MDP that specifies action $a$ that is taken in state $s$. 
At each time step, an agent observes a state $s \in \cS$ and chooses an action $a \in \cA$. After each time step, the agent gets an immediate reward $r$ and next state $s^{\prime}\in\cS$ in return for the action taken. In this paper, the policy is assumed to be a deterministic function, denoted by $a=\mu(s)$.

DPG algorithm can be used to handle multi-dimensional continuous actions in many continuous control problems \cite{Lillicrap:16}. 
DPG utilizes a novel actor-critic architecture that consists of two components, namely, actor and critic \cite{Sutton:17}.
The actor learns to produce a deterministic policy based on the state, while the critic learns to estimate the true action-value (Q-value) function of an action given a state. The policy and the value function are parameterized by neural networks as $\mu_{\phi}$ and $\bQ_{\theta}^{\mu}$, respectively. 
The critic estimates the policy gradient from the learned $\bQ_{\theta}^{\mu}$ and sends it to the actor to update the policy $\mu_{\phi}$ at the same time. 
As a deep variant of DPG, DDPG combines deep neural networks with the actor-critic architecture \cite{Lillicrap:16}.

MA-DDPG is increasingly used within a diverse range of applications involving multi-agent environments with multi-dimensional continuous action space. 
There are basically two design factors to consider: allowing agents to share a single critic and augmenting a critic with the policies of other agents.
By combining features of the original designs in \cite{Ryan:17} and \cite{Foerster:18}, in this paper we consider a MA-DDPG framework where
one critic is shared by all agents and the centralized critic is augmented with policies of the agents.

\subsection{MA-DDPG in MISO IFC }
\begin{figure}
\centerline{\includegraphics[trim=0cm 0.4cm 0cm 0.4cm,clip,width=0.99 \linewidth]{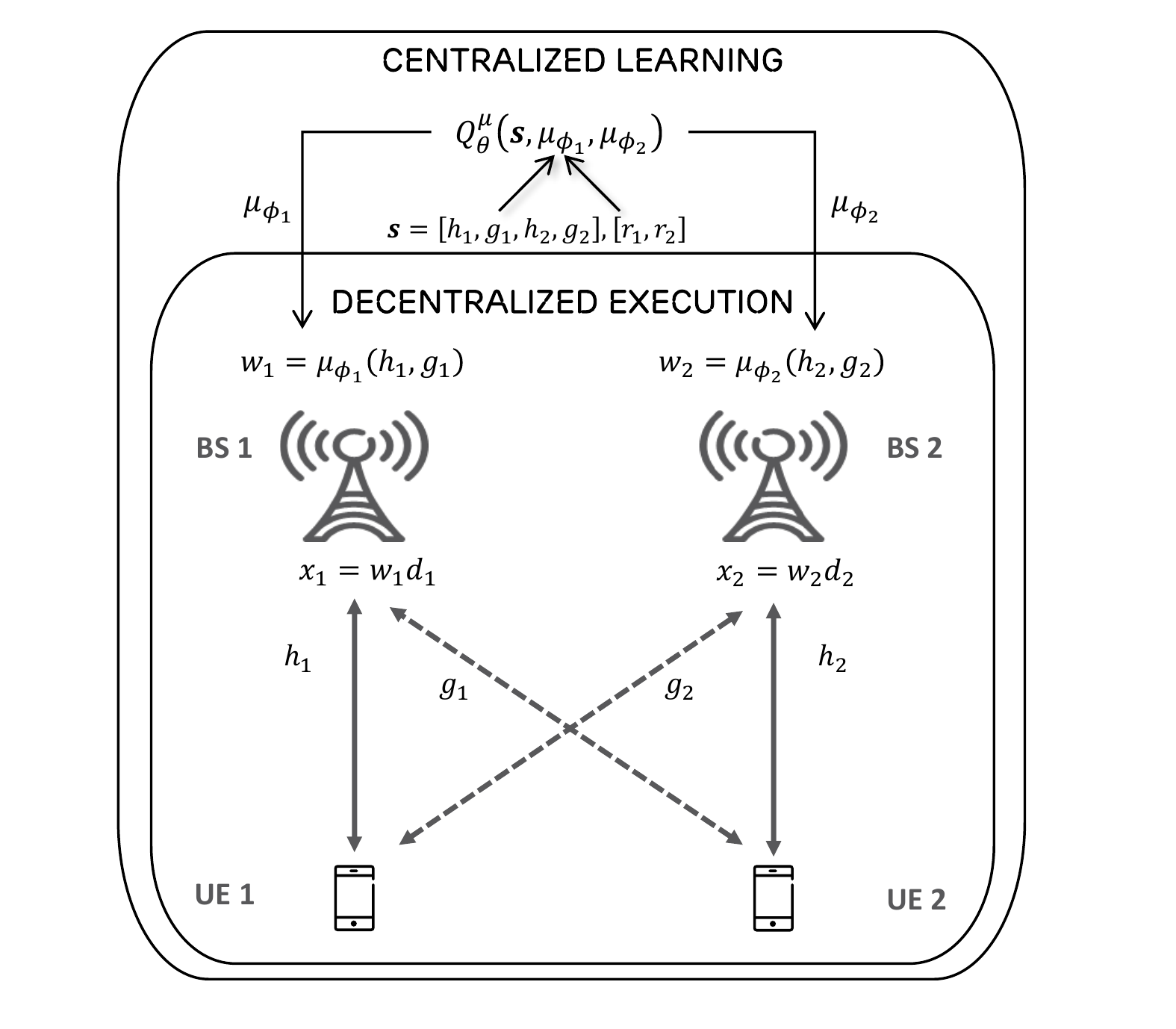}}
\caption{MA-DDPG in MISO IFC}
\label{fmaddpg}
\end{figure}

Figure \ref{fmaddpg} illustrates MA-DDPG model for the MISO IFC setup under the following assumptions:
\begin{itemize}
\item Non-real time communication available for the critic to learn an action-value function in a centralized manner based on global information about environmental channel states and policies of both agents.
\item To fulfill the real or near-real time requirement of precoding scheme, each agent $i$ at the $i$-th BS chooses a precoding vector $\vw_i$ based on local information given by the partial channel observation $\vh_i$ and $\vg_i$ only.
\end{itemize}

As shown in Figure \ref{fmaddpg}, the MA-DDPG extends the actor-critic policy gradient method to provide a framework of centralized training with decentralized execution, improving training stability at training phase and performance robustness at execution phase. 
More importantly, the framework allows non-real time learning based on global information and real or near-real time execution based on local information, therefore making it practical for real-world cellular environments.

In TDD-based systems, downlink CSI can be derived from uplink channel observations thanks to channel reciprocity.
For instance, in current LTE and NR specifications, each BS $i$ can estimate downlink channel state of $\vh_i$ and $\vg_i$ based on uplink sounding reference signal (SRS) transmitted by the desired and interfering UE.
Therefore, the partial state $\vs_i$ (or partial observation $\vo_i$ more specifically) at BS $i$ can be defined, directly from the local observation of $\vh_i$ and $\vg_i$, as 
\bea \label{eq:state}
\vs_i=[\vh_i,\vg_i].
\eea

The actor $i$ chooses an action for a given state $\vs_i$ by using a deterministic policy 
\bea 
\va_i=\mu_{\phi_i}\left(\vs_i\right),
\eea
where the output action $\va_i=[a_{i,1},a_{i,2},\cdots,a_{i,n_t}]^T$ is used to determine a precoding vector $\vw_i=[w_{i,1},w_{i,2},\cdots,w_{i,n_t}]^T$.

While we use a deterministic policy $\mu_{\phi_i}$ that always yields the same action for the same state, a stochastic policy is desirable for exploration at training time. DDPG uses a stochastic behavior policy to select actions, different from the learned policy.
In order to ensure sufficient exploration, in this paper, we perturb the deterministic action $a=\mu_{\phi_i}(s)$ by adding a noise vector
whose entries are i.i.d. according to $\cN_{\dC}(\mo, \sigma_p^2\mI)$, as described in \cite{Lillicrap:16}. 

Each agent $i$ receives as a reward the rate $r_i$ given by \eqref{eq:achi} as a function of the environmental state in \eqref{eq:state} and actions taken by the stochastic behavior policy. 
In order to achieve a rate pair on the pareto-boundary of the achievable rate region, 
the agents should behave in a cooperative manner to maximize the collective reward. 
The collective reward $r_{c,\alpha}$ for achieving a pareto-optimal precoding strategy can be defined as a weighted sum of achieved rates $ r_1 $ and $ r_2$ \cite{Park:2010}
\bea
r_{c,\alpha}=\alpha\cdot r_1 + (1-\alpha)r_2
\eea
where $\alpha \in [0,1]$ denotes the weighting scalar between the two rewards $r_1$ and $r_2$.

We note that the weighting factor $\alpha$ determines which pareto-optimal rate pair to achieve by specifying a straight line with slope $-\frac{\alpha}{(1-\alpha)}$ to the pareto-boundary. 
In the simulation section, we will consider three different values, $\alpha=1/2, 2/3,$ and $3/4$, which will achieve the point of rate pairs on the pareto-boundary where the slope of tangent is equal to $-1, -2$ and $-3$, respectively. 

The centralized critic aims to maximize its total expected return $R_c=\sum_{t=0}^{\infty}\gamma^{t}r_{c,\alpha}^t$, where $\gamma\in\left[0,1\right]$ denotes a discounting factor to the future rewards $r_{c,\alpha}^t$ for $t=1,2,\cdots,$ relative to the immediate reward $r_{c,\alpha}^0$.
To this end, the true action-value function is approximated by the critic network
\bea \label{eq:critic}
\bQ_{\theta}^{\mu}\left(\vs,\va_1,\va_2\right)
\eea
where $\mu=[\mu_{\phi_1},\mu_{\phi_2}]$ and the global state $\vs$ is given by $\vs=[\vs_1,\vs_2]$.

Equation \eqref{eq:critic} shows that the critic function uses the policies of both agents so that each agent can learn approximate models of other agents from the learned critic. 

At each time step, according to Q-learning, the critic updates the value-function parameters ${\theta}$ as follows:
\be \label{eq:cupdate}
\theta \leftarrow \theta+\eta_c\left(Y^{\theta} - \bQ_{\theta}^{\mu}\left(\vs,\va_1^p,\va_2^p\right)\right)
\triangledown_{\theta} \bQ_{\theta}^{\mu}\left(\vs,\va_1^p,\va_2^p\right),
\ee
where $\eta_c\in\left[0,1\right]$ is a critic learning rate, $\triangledown_{\theta} \bQ_{\theta}^{\mu}\left(\cdot\right)$ denotes the vector of partial derivatives with respect to the components of $\theta$, $\va_i^p$ is the noisy version of $\va_i=\mu_{\phi_i}(\vs_{i})$, and $Y^{\theta}$ indicates the newly estimated value on the current step, assuming the deterministic actions on the next step $\vs^{\prime}=[\vs_1^{\prime},\vs_2^{\prime}]$, which is given by 
\be 
Y^{\theta}=r_{c,\alpha} + \gamma \bQ^{\mu} _{\theta}(\vs^{\prime}, \mu_{\phi_1}\left(\vs_1^{\prime}\right),\mu_{\phi_2}\left(\vs_2^{\prime}\right) ) ).
\ee 

The actor $i$ estimates the policy parameters $\phi_i$ that maximize the expected reward by updating the parameters via a gradient ascent
\be
\phi_i \leftarrow \phi_i+\eta_a \triangledown_{\phi_i} J(\mu_{\phi_i}),
\ee
where $\eta_a$ is an actor learning rate and, according to the DPG algorithm, the gradient $ \triangledown_{\phi_i} J(\mu_{\phi_i})$ is obtained by
\be \label{eq:grad}
\triangledown_{\phi_i} J(\mu_{\phi_i}) = \triangledown_{\phi_i}\mu_{\phi_i}(\vs_i) \triangledown_{a_i}\bQ_{\theta}^{\mu}\left(\vs,a_1,a_2\right)|_{a_i=\mu_{\phi_i}(\vs_{i})}.
\ee

After training is completed, the local actors can execute the learned policies only based on the local CSI at execution phase, successfully addressing the challenge of multiple agents with partial observability. 

\subsection{Phase ambiguity elimination}
In this subsequent section, 
we address the phase ambiguity issue in the commonly used vector (or matrix) representation of wireless channel states 
and then present a method to improve MA-DDPG training in MISO IFC. 
\newline
In radio communications, a passband channel state is represented by a complex-valued baseband equivalent. 
In our signal model given in \eqref{eq:rxsig1} and \eqref{eq:rxsig2}, the channel vectors $\vh_1$, $\vg_1$, $\vh_2$ and $\vg_2$ 
can be expressed as a complex-valued representation with respect to amplitude and phase, denoted by  
\bea \label{eq:rep}
\vc=[a_1\exp(j\vartheta_1),a_2\exp(j\vartheta_2),\cdots,a_{n_t}\exp(j\vartheta_{n_t})],
\eea
where $a_i$ and $\vartheta_i$ are the amplitude and phase of the $i$-th element of vector $\vc$.
\newline
We note that the channel state given by $\vc$ in \eqref{eq:rep} has inherent phase ambiguity resulting from the complex-valued baseband signal representation.
More specifically, all the phase-shifted states  $\exp(j\varphi)\vc$ with arbitrary phases $\varphi$ are supposed to lead to the same action as that of the original state $\vc$. 
From the wireless system design point of view, this phase ambiguity should not be a problem, 
but it introduces a many-to-one mapping issue between a set of phase-shifted states with different offsets and one target optimal action, which further complicates the training task, and thereby, degrades the performance in a multi-agent learning system.
In order to combat the degradation due to the many-to-one mapping nature of state-action pairs, we propose a PAE method as a pre-processing on channel states $\vh_1$, $\vg_1$, $\vh_2$ and $\vg_2$, that maps  channel states with phase ambiguity into the same state. 
The training method can utilize any mapping function $f_{PAE}(\cdot)$ that eliminates inherent phase ambiguity. For instance, in this paper we consider a PAE mapping function that maps each state $\vc $ onto one state whose first element is purely real-valued as follows: 
\bea
f_{PAE}(\vc)=[a_1,a_2\exp\left(j\left(\vartheta_2-\vartheta_1\right)\right),\cdots, \nonumber \\ 
a_{n_t}\!\exp\left(j\left(\vartheta_{n_t}-\vartheta_1\right)\right)].
\eea
\newline
In summary, the final state representation can be obtained by applying the mapping function $f_{PAE}$ to $\vh_i$ and $\vg_i$ as
\bea \label{eq:statenew}
\vs_i^{PAE}=[f_{PAE}(\vh_i),f_{PAE}(\vg_i)].
\eea
\newline
In the following section, we will show that the proposed PAE training method can achieve a faster convergence and a better performance compared to the trivial approach given in \eqref{eq:state} without consideration of phase ambiguity elimination.

\section{Numerical results} \label{sec:result}
In this section, we provide simulation results and comparisons with the pareto-boundary to demonstrate the optimality of MA-DDPG framework in MISO IFC setup.
We consider BSs with three transmit antennas, i.e., $n_t=3$, and fix the average SNR to be 10 dB in all the simulations. 
Note that the all the precoding weight vectors given by the numerical methods and learned by MA-DDPG should be normalized to make $||\vw_i||=1$. We implemented the MA-DDPG model in TensorFlow 2, training a critic and two actors all with three hidden fully-connected layers.
Each episode length is defined to have 10,000 time steps. We start from a perturbation variance $\sigma_p^2=0.1$ and multiply it by a decaying factor of 0.993 over episodes.

\begin{figure}
\centerline{\includegraphics[trim=0cm 0.3cm 0cm 0.3cm,width=0.94 \linewidth]{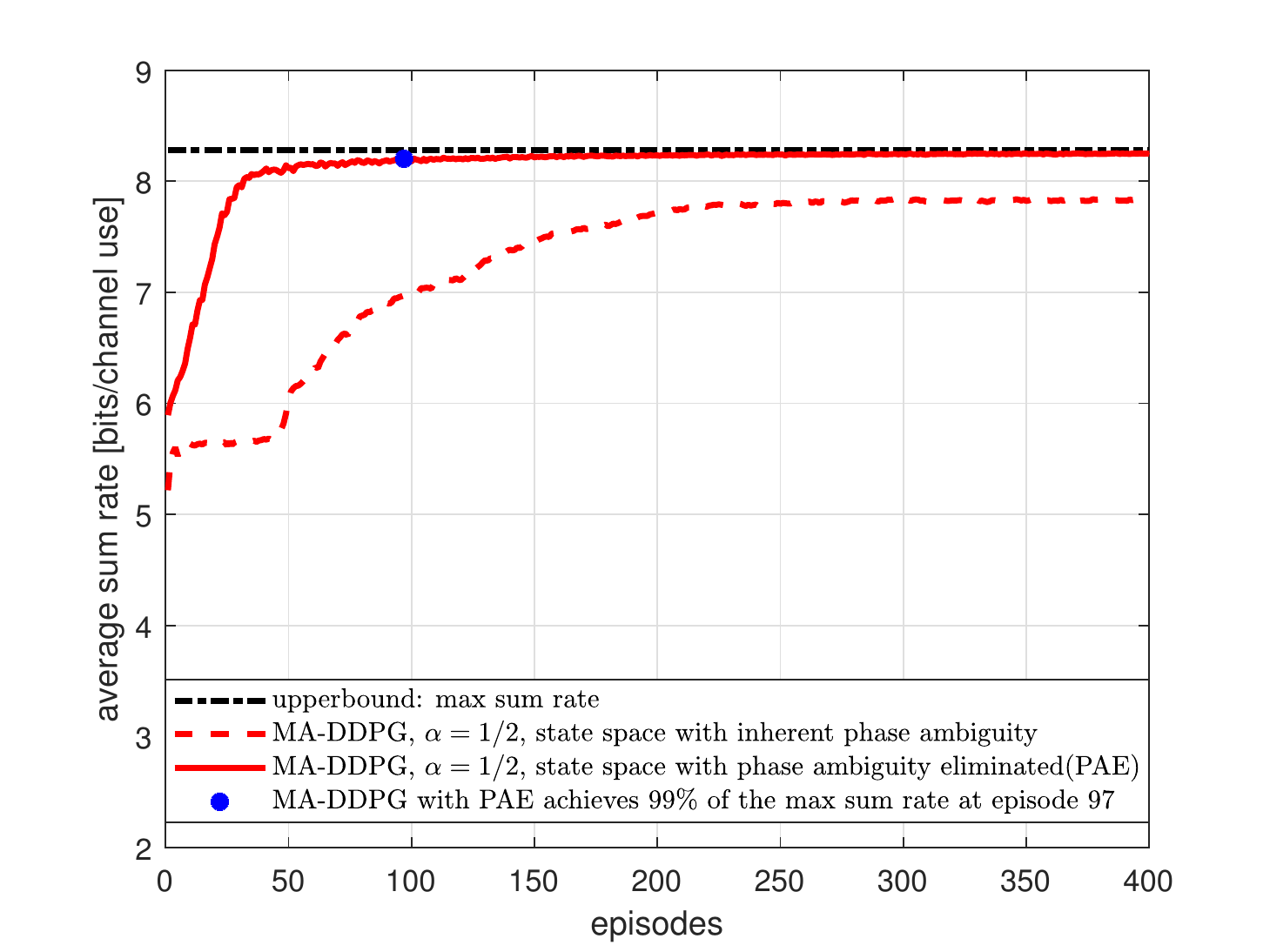}}
\caption{Training convergence and performance impact due to phase ambiguity}
\label{fpae}
\end{figure}

Figure \ref{fpae} illustrates the impact of phase ambiguity on learning performance, comparing with the theoretical upper-bound of MA-DDPG with $\alpha=1/2$.
After each episode, consisting of 10,000 training steps, we evaluate the learned policies in terms of average sum rate using a test set of 5000 unseen samples.
MA-DDPG with PAE achieves more than $99\%$ of the maximum achievable sum rate after 97 episodes while only a maximum of $94\%$ is achieved without PAE due to the many-to-one mapping nature of state-action pairs.
The simulation results show that the proposed PAE training method can achieve a faster and even better learning curve in a MISO IFC environment.

\begin{figure}
\centerline{\includegraphics[trim=0cm 0.3cm 0cm 0.3cm,width=0.94 \linewidth]{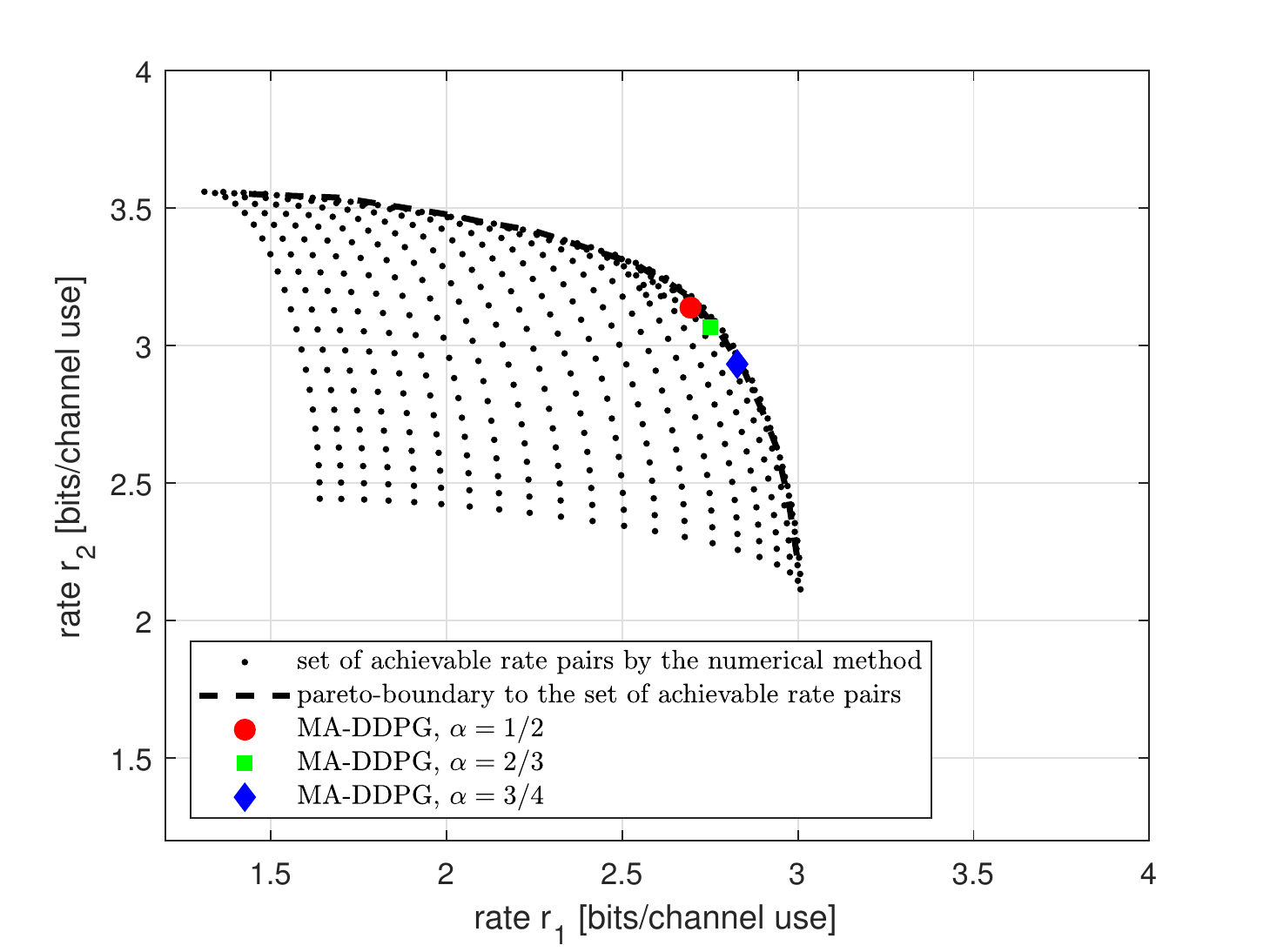}}
\caption{Achieved rate pairs by MA-DDPG with $\alpha=1/2, 2/3,$ and $3/4$}
\label{fsample}
\end{figure}

\begin{figure}
\centerline{\includegraphics[trim=0cm 0.3cm 0cm 0.3cm,width=0.94 \linewidth]{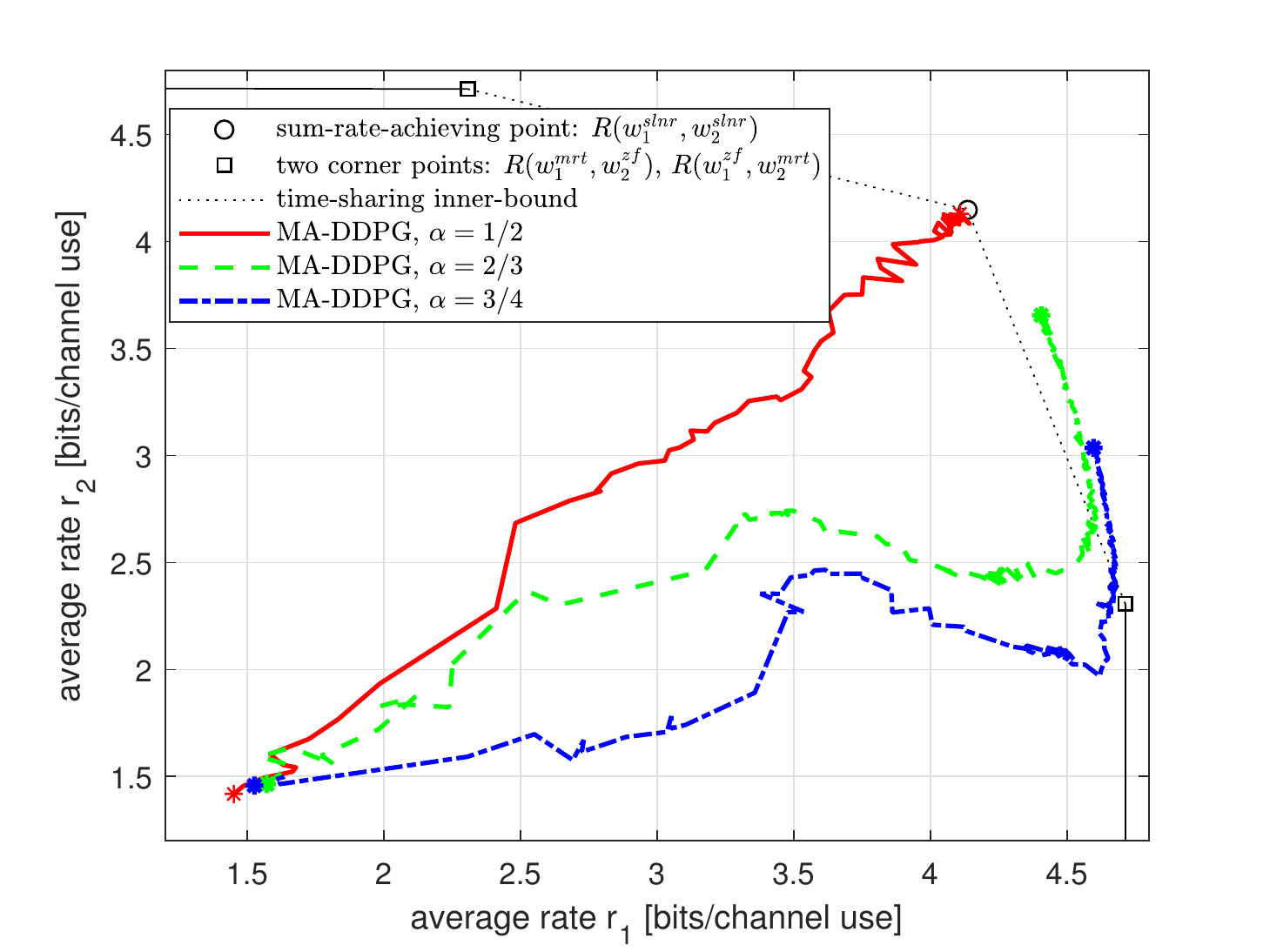}}
\caption{Learning behaviors of MA-DDPGs in terms of the average rate pairs}
\label{frateregion}
\end{figure}

In the following figures, we provide numerical results to demonstrate the optimality of MA-DDPG with the proposed PAE method. 
MA-DDPG models are trained over 200 episodes and quantitative comparisons are provided in terms of achieved rate pair for a random test sample or in terms of average rate pair for 5000 test samples. 
For a given sample $\vh_1=[-0.569+j0.227, -0.018+j0.456, -0.213+j0.254]$,
$\vg_1=[-0.054-j0.240, 0.298 - j0.232, 0.334 - j0.403]$,
$\vh_2=[-0.846 - j0.287, -0.129 + j0.073, -0.098 + j0.499]$, and
$\vg_2=[0.636 - j0.493, -0.167 - j0.050, 0.204 + j0.460]$, the achieved rate pairs by MA-DDPGs are shown in Figure \ref{fsample} in comparison to the achievable rate pairs by the numerical method.
Figure \ref{frateregion} shows learning curves of MA-DDPGs over 200 episodes in comparison to the pareto-optimal rate pairs, namely, the sum-rate optimal reference point $R(\vw_1^{slnr},\vw_2^{slnr})$ as well as the two corner points $R(\vw_1^{mrt},\vw_2^{zf})$ and $R(\vw_1^{zf},\vw_2^{mrt})$, based on 5000 test samples that haven't seen by the agents before.
These simulation results demonstrate that the decentralized actors with partial observability are able to discover optimal coordination strategies with the aid of the centralized critic in MISO IFC environments.


\section{Conclusion} \label{sec:conc}
In this paper, we have proposed a multi-agent deep reinforcement learning approach for precoding
method in multi-cell multi-user MIMO systems.
We have demonstrated that a MA-DDPG framework is able to automatically learn a near-optimal precoding policy in MISO IFC.
In particular, we have shown that the MA-DDPG framework allows for centralized learning with decentralized execution at different levels of observability and time requirement, which is a practical approach for real-world cellular environments.
Furthermore, we have addressed the phase ambiguity issue with the conventional baseband signal representation used in radio communications and proposed the phase 
ambiguity elimination method. The numerical simulation results show that the phase-ambiguity elimination in state space is crucial for successful training of MADRL in wireless communication systems. The proposed method can be also applied in precoding action space.

\bibliographystyle{ieeetr}
\bibliography{REFbib}

\end{document}